\documentclass{jfm}
\usepackage{graphicx}
\usepackage{epstopdf, epsfig}

\usepackage{xcolor}
\usepackage{amsmath}
\usepackage{tikz}
\definecolor{myblue}{rgb}{0.12156862745098039, 0.4666666666666667, 0.7058823529411765}

\definecolor{myorange}{rgb}{1.0, 0.4980392156862745, 0.054901960784313725}

\definecolor{mygreen}{rgb}{0.17254901960784313, 0.6274509803921569, 0.17254901960784313}

\definecolor{myred}{rgb}{0.8392156862745098, 0.15294117647058825, 0.1568627450980392}

\definecolor{mypink}{rgb}{0.8901960784313725, 0.4666666666666667, 0.7607843137254902}

\newcommand{\blueline}{\raisebox{2pt}{\tikz{\draw[-,myblue,solid,line width = 1.25pt](0,0) -- (5mm,0);}}}

\newcommand{\orangeline}{\raisebox{2pt}{\tikz{\draw[-,myorange,solid,line width = 1.25pt](0,0) -- (5mm,0);}}}

\newcommand{\greenline}{\raisebox{2pt}{\tikz{\draw[-,mygreen,solid,line width = 1.25pt](0,0) -- (5mm,0);}}}

\newcommand{\pinkline}{\raisebox{2pt}{\tikz{\draw[-,mypink,solid,line width = 1.25pt](0,0) -- (5mm,0);}}}

\newcommand{\redline}{\raisebox{2pt}{\tikz{\draw[-,myred,solid,line width = 1.25pt](0,0) -- (5mm,0);}}}

\newcommand{\blackline}{\raisebox{2pt}{\tikz{\draw[-,black,solid,line width = 1.25pt](0,0) -- (5mm,0);}}}

\newcommand{\blackdashedline}{\raisebox{2pt}{\tikz{\draw[-,black,densely dashed,line width = 1.25pt](0,0) -- (5mm,0);}}}

\newcommand{\blackdottedline}{\raisebox{2pt}{\tikz{\draw[-,black,dotted,line width = 1.25pt](0,0) -- (5mm,0);}}}

\shorttitle{Significance of the strain-dominated region around a vortex }
\shortauthor{K. Menon and R. Mittal}

\title{Significance of the strain-dominated region around a vortex on induced aerodynamic loads}

\author{Karthik Menon\aff{1}
  \corresp{\email{kmenon@jhu.edu}}
 \and Rajat Mittal\aff{1}
 \corresp{\email{mittal@jhu.edu}}}

\affiliation{\aff{1}Department of Mechanical Engineering, Johns Hopkins University, Baltimore, MD 21218, USA}

\begin{document}

\maketitle

\begin{abstract}
The ability of vortices to induce aerodynamic loads on proximal surfaces plays a significant role in a wide variety of flows. However, most studies of vortex-induced effects primarily focus on analyzing the influence of the rotation-dominated cores of vortices. In this work, we show that not only are vortices in viscous flows surrounded by strain-dominated regions, but that these regions are dynamically important and can sometimes even dictate the induced aerodynamic loads. We demonstrate this for a pitching airfoil, which exhibits dynamic stall and generates several force-inducing vortices. Using a data-driven force partitioning method, we quantify the influence of vortices as well as vortex-associated strain to show that our current understanding of vortex-dominated phenomena, such as dynamic stall, is incomplete without considering the substantial effect of strain-dominated regions that are associated with vortices.
\end{abstract}


\section{Introduction}
\label{sec:intro}

Vorticity play a singularly important role in fluid dynamics. This is because (a) the generation/appearance of vorticity in a flow provides direct information regarding the influence of viscosity; (b) flow instabilities and transition to turbulence occur in vorticity dominant regions such as shear layers, boundary layers and vortices; and finally, (c) vorticity organized in the form of vortices is known to induce aerodynamic loads on proximal surfaces, 

The current paper concerns this third aspect of vorticity i.e. the ability of vortices to induce aerodynamic loads on immersed surfaces -- a fact that has been exploited to great effect in nature \citep{Ellington1996Leading-edgeFlight,Triantafyllou2000HydrodynamicsSwimming} as well as in engineered wings and control surfaces \citep{Carr1988ProgressStall,Eldredge2019}. Indeed, unsteady aerodynamics is one arena where a number of studies have focused on determining the aerodynamic loads induced by vortices in a flow \citep{Dickinson1993UnsteadyNumbers,Sane2003TheFlight,Shyy2007FlappingVortices,Chen2010ThePlate,PittFord2013LiftVortex,Xia2013LiftPlate,Mulleners2013DynamicDevelopment,Gharali2013DynamicVelocity}. This task is made difficult not only by the fact that most flows consist of multiple interacting vortices and shear layers, but also that there continues to be significant ambiguity regarding the very definition of a vortex \citep{Haller2015LagrangianStructures} and its relationship to the induced pressure field. 

One method for vortex identification that is commonly used is the so called ``$Q$-criterion" \citep{Hunt1988EddiesFlows,Jeong1995OnVortex}. The quantity $Q$ is the second invariant of the velocity gradient tensor and can be defined  as  
$Q = \frac{1}{2}\left(||\bar{\Omega}||^2 - ||\bar{S}||^2 \right)$. Here $\bar{S}$ and $\bar{\Omega}$ are the strain-rate and rotation tensors given by $\bar{S} = \frac{1}{2} \left[ \boldsymbol{\nabla} \boldsymbol{u} + \left(\boldsymbol{\nabla} \boldsymbol{u} \right) ^T \right]$ and  $\bar{\Omega} = \frac{1}{2} \left[ \boldsymbol{\nabla} \boldsymbol{u} - \left(\boldsymbol{\nabla} \boldsymbol{u} \right) ^T \right]$ respectively. A vortex, according to the $Q$-criterion, is then defined as a connected region where $Q>0$. This corresponds to regions where the Frobenius norm of the rotation tensor is greater than that of the strain-rate tensor.

Interestingly, while the quantity $Q$ has mostly been associated with the flow kinematics, \cite{Jeong1995OnVortex} noted that the pressure Poisson equation for incompressible flow, derived by taking the divergence of the Navier-Stokes equation, can be expressed as: 
\begin{equation}
    \nabla^2p = -\rho \boldsymbol{\nabla} \cdot \left( \boldsymbol{u} \cdot \boldsymbol{\nabla}  \boldsymbol{u} \right) 
    \equiv \rho \boldsymbol{\nabla}  \boldsymbol{u} : \left( \boldsymbol{\nabla}  \boldsymbol{u} \right)^T
    \equiv 2 \rho Q. 
    \label{eq:ppe}
\end{equation}
Equation \ref{eq:ppe} shows that $Q$ appears in the source term of the pressure Poisson equation and is thus, not merely a kinematic quantity, but one with direct implications for the \emph{dynamics} of the flow as well. Furthermore, while regions of positive $Q$, which are used to identify vortices, may draw much of our attention, equation \ref{eq:ppe} shows that regions with negative $Q$, i.e. strain dominated regions, are no less significant with regard to their effect on pressure. 

A question that immediately arises from equation \ref{eq:ppe} is: do vortices have regions of negative $Q$ associated with them and, if so, how significant are these regions? To answer this, we first note that the following Neumann boundary condition completes the prescription of the above boundary value problem: 
$\boldsymbol{\hat{n}} \cdot \boldsymbol{\nabla} p =   \boldsymbol{\hat{n}} \cdot \left[ -\rho{ \frac{d \boldsymbol{u}}{dt}} + \mu \nabla^2 \boldsymbol{u}  \right]$.
We now consider a vortex in an infinite quiescent flow for which, the boundary condition at the far-field would be homogeneous. The divergence theorem applied to the pressure Poisson equation would therefore require that $\int_{V_f} Q  \ dV \equiv 0$, where $V_f$ is the volume of the flow domain. This implies that for such a vortex, the volume integral of the negative $Q$ region is equal to the volume integral of positive $Q$ in the vortex core. Furthermore, immediately outside the region of positive $Q$, the velocity gradients have to be non-zero and therefore, this region immediately surrounding the vortex core should have $Q < 0$. Thus, for any vortex in a viscous flow, the positive-$Q$ core is necessarily associated with a negative-$Q$ ``corona" of equivalent significance. This can easily be confirmed for exact vortex solutions of the Navier-Stokes equations such as the Rankine vortex \citep{acheson1990elementary}, the Lamb-Oseen vortex \citep{lamb1993hydrodynamics}, and others. While the presence of boundaries and other flow complexities might eliminate the strict equipartitioning of $Q$ for a vortex, we nevertheless expect that every vortex will have a significant corona of negative $Q$ around it. This observation, combined with equation \ref{eq:ppe}, suggests that the strain dominated region around a vortex could have a significant contribution to the net aerodynamic load induced by a vortex. However, although the overall effect of these strain-dominated regions on aerodynamic loading has implicitly been included in past studies, our physical intuition regarding aerodynamic loading has generally been tied to the growth, motion, and overall evolution of the rotational core of vortices. The role of the strain-induced region on force production has, thus far, not garnered significant attention.  

In this paper, we examine the aerodynamic loads induced by vortices on an immersed surface, and in particular, we focus on the distinct roles played by regions of positive as well as negative $Q$ associated with vortices. For this analysis, we choose to focus on the problem of a sinusoidally pitching airfoil that exhibits dynamic stall, and generates leading-edge (LEV) and trailing-edge vortices (TEVs). We employ a force partitioning approach that allows us to precisely quantify the force contributions of various vortices and vortical regions in the flow. Using this analysis, we examine the phenomenon of dynamic stall and show that our understanding of vortex-induced loads is incomplete without also discussing the role played by the corona of negative $Q$ around a vortex. 

\section{Methods and problem description}
\subsection{Computational model and numerical method}
\label{sec:num_meth}
The two-dimensional computational model studied in this work consists of a sinusoidally pitching NACA0015 airfoil with a rounded trailing edge, immersed in a uniform free-stream flow with velocity $U_{\infty}$. The flow simulations have been performed using the sharp-interface immersed boundary method of \cite{Mittal2008ABoundaries} and \cite{Seo2011AOscillations}. This method is well-suited for moving boundary problems as it allows the use of simple non-conformal Cartesian grids to simulate a variety of different shapes and motions of the immersed body. Further, the ability to preserve the sharp-interface around the immersed boundary ensures very accurate computations of surface quantities. The Navier-Stokes equations are solved using a fractional-step method. Spatial derivatives are discretized using second-order central differences, and time-stepping is achieved using the second-order Adams-Bashforth method. The pressure Poisson equation is solved using a geometric multigrid method. This code has been extensively validated in previous studies for a wide variety of stationary as well as moving boundary problems \citep{Mittal2008ABoundaries,Seo2011AOscillations}, where its ability to maintain local (near the immersed body) as well as global second-order accuracy has been established.

The Reynolds number of the flow is defined as $Re = U_{\infty} C /\nu$, where $C$ is the chord-length of the airfoil and $\nu$ is the kinematic viscosity of the fluid. The pitch oscillation is prescribed as: $\theta = \theta_0 + A_{\theta}\sin{(2 \pi f t)}$ where $\theta$ is the instantaneous pitch angle, $\theta_0$ is the mean pitch angle and $A_{\theta}$ is the pitch amplitude. The dimensionless pitching frequency is given by $f = F C/U_{\infty}$ where $F$ is the pitch frequency and $t$ is the dimensionless time (also scaled by $U_{\infty}$ and $C$). This model is placed symmetrically in a large $21C \times 23C$ computational domain with a grid of size $384\times320$ points. The grid around the airfoil is isotropic, and corresponds to about $125$ points along the chord. A Dirichlet velocity boundary condition is used at the inlet boundary of the domain, and zero-gradient Neumann conditions are specified at all other boundaries. Grid refinement studies and other details can be found in \cite{Menon2019,Menon2020AerodynamicNumbers}. 

\subsection{Force partitioning and vortex-induced aerodynamic loads}
\label{sec:fpm}
\newcommand{\fpmkin}{C^{\kappa}_{F_i}}
\newcommand{\fpmvif}{C^{\omega}_{F_i}}
\newcommand{\fpmshr}{C^{\sigma}_{F_i}}
\newcommand{\fpmpot}{C^{\Phi}_{F_i}}
\newcommand{\fpmext}{C^{\Sigma}_{F_i}}
\newcommand{\vol}{{V_f}}
\begin{figure}
  \centerline{\includegraphics[scale=0.8]{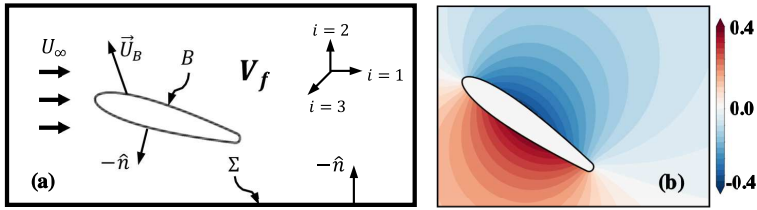}}
  \caption{(a) Schematic of the setup for the force partitioning method (FPM), along with relevant symbols. (b) Sample snapshot of the $\phi_2$ field around an airfoil.}
\label{fig:fpm_schematic}
\end{figure}
A key element of the current study is the ability to estimate the aerodynamic loads induced by vortices. This is accomplished using the force partitioning method (FPM) of \cite{Zhang2015CentripetalInsects} and \cite{Menon2020QuantitativeApproach}, a concise description of which is presented here. 
The first step in the FPM is the calculation of an auxiliary potential field, $\phi_i$, which satisfies the following equation: 
\begin{equation}
  {\nabla}^2 \phi_{i} = 0, \ \ \mathrm{ in } \ \ V_f \ \ \mathrm{ with } \ \
  \boldsymbol{\hat{n}} \cdot \boldsymbol{\nabla} \phi_{i}=
    \begin{cases}
      n_i \;, \; \mathrm{on} \; B \\
      0 \; \;, \; \mathrm{on} \; \Sigma 
    \end{cases}
    \ \ \ \mathrm{ for} \  i=1, 2, 3.
  \label{eq:fpm_potential}
\end{equation}
Here, as shown in figure \ref{fig:fpm_schematic}(a), $V_f$ is the fluid domain with outer boundary $\Sigma$ and $B$ is the surface of the immersed body. The unit normal pointing outward from the fluid domain and into the internal and external boundaries, $B$ and $\Sigma$, is given by $\hat{n}$ and the index $i$ determines the vector component of the aerodynamic force that is being partitioned. In the current two-dimensional problem, $i=1$ and $2$ correspond to the drag and lift forces respectively on the airfoil. We note that this auxiliary potential is a function only of the instantaneous position and shape of the immersed body and the outer domain boundary. Figure \ref{fig:fpm_schematic}(b) shows an example $\phi_2$ field around an airfoil. 

The next step is to project the Navier-Stokes equation onto the gradient of the auxiliary potential and integrate over the volume of the fluid domain. This results in, 
\begin{equation}
  -\int_{V_f} \boldsymbol{\nabla} p \cdot \boldsymbol{\nabla} \phi_i \  dV =  \int_{V_f} \rho \left[ \frac{\partial \boldsymbol{u}}{\partial t}  +  \boldsymbol{u} \cdot \boldsymbol{\nabla} \boldsymbol{u} 
  -  \nu   \nabla^2 \boldsymbol{u}  
  \right]  \cdot \boldsymbol{\nabla} \phi_i \ dV \ \mathrm{ for} \  i=1, 2, 3
  \label{eq:projected_NS}
\end{equation}
By making use of properties of $\phi_i$ as prescribed in equation \ref{eq:fpm_potential}, as well as the incompressibility constraint and the divergence theorem, the above expression can be rewritten for the pressure-induced force on the body as follows:
\begin{equation}
\begin{aligned}
F_i =
  \int_{B}  p \ n_i \  dS  =  
  &- \rho \int_{B} \boldsymbol{\hat{n}}  \cdot \left( \frac{d  \boldsymbol{U}_B}{d t} \phi_i \right) \ dS + \rho \int_{V_f} \boldsymbol{\nabla} \cdot \left( \boldsymbol{u} \cdot \boldsymbol{\nabla}  \boldsymbol{u} \right) \phi_i \ dV \\
  &- \mu \int_{V_f}  \left( \boldsymbol{\nabla}^2 \boldsymbol{u} \right)  
  \cdot \boldsymbol{\nabla} \phi_i \  dV - \rho \int_{\Sigma} \boldsymbol{\hat{n}}  \cdot \left( \frac{d  \boldsymbol{u}}{d t} \phi_i \right) \ dS
\ \  \mathrm{ for} \  i=1 , 2 , 3
\end{aligned}
  \label{eq:FPM}
\end{equation}
where $\boldsymbol{U}_B$ is the local velocity of the immersed surface. The four terms on the right-hand-side of equation \ref{eq:FPM} represent distinct components of the pressure-induced force on the surface. The first term contains the well-known linear acceleration reaction (i.e. added mass) force \citep{Menon2020OnPartitioning} as well as the centripetal acceleration reaction, which was shown by \cite{Zhang2015CentripetalInsects} to be an important lift generation mechanism for insect wings. The third term is the pressure force associated with viscous diffusion in the flow and this term is usually significant only for low-Reynolds number flows. The fourth term is associated with flow acceleration at the outer boundary and this term is negligible for large domains and steady freestream flows. The surface shear stress also induces a force on the surface that is independent of the pressure-induced force, but this component of the force is also very small for high Reynolds number flows. Further details of this FPM applied to aerodynamic forces as well as moments can be found in \cite{Zhang2015CentripetalInsects} and \cite{Menon2020OnPartitioning,Menon2020QuantitativeApproach}.

The second term in the above expression, which is the primary focus of the current study, can be expressed as, 
\begin{equation}
 F_i^\omega = \rho \int_{V_f}  \boldsymbol{\nabla} \cdot \left( \boldsymbol{u} \cdot \boldsymbol{\nabla}  \boldsymbol{u} \right) \phi_i \ dV \equiv - \rho \int_{V_f} 2 \ Q \ \phi_i \ dV\ \ \  \mathrm{ for} \  i=1 , 2 , 3 .
\label{eq:VIF}
\end{equation}
Given the centrality of $Q$ in this force component, we identify this term as the ``vortex-induced force" and denote it as $\boldsymbol{F}^\omega$. This vortex-induced force is found to be a dominant component in a variety of flows including those over flapping wings \citep{Zhang2015CentripetalInsects}, bluff bodies \citep{Menon2020OnPartitioning}, pitching foils \citep{Menon2020QuantitativeApproach} and delta wings \citep{Li2020VortexWing}. This is also true for the particular pitching airfoil problem relevant to this paper. As we will show in the next section, the order of magnitude of vortex-induced force oscillations is $O(1)$ and this term accounts for the majority of the total lift on the airfoil. In comparison, the viscous and added-mass terms show oscillations with orders of magnitude $O(10^{-1})$ and $O(10^{-2})$ respectively (not shown here for brevity). Plots of these force contributions for sample pitching airfoil cases can be found in \cite{Menon2020QuantitativeApproach}.

It is also useful to note that the volume-integral form of the vortex-induced force term in equation \ref{eq:VIF} allows us to compute the force contribution of any region (or flow structure) of interest by constructing appropriate spatial volumes enclosing the flow structure, and evaluating the integral over this volume. \cite{Zhang2015CentripetalInsects} used this idea to determine the contribution to lift of leading-edge vortices over flapping insect wings and \cite{Menon2020OnPartitioning} employed this to show that during stationary-state, transverse flow-induced vibration of a bluff body, the wake vortices serve to damp the vibrations whereas the boundary layer on the oscillating cylinder surface tends to promote oscillations.

The construction of such integral volumes in complex, time-varying flows is itself a challenge and in this work, we perform this task using a novel automated data-driven procedure described in \cite{Menon2020QuantitativeApproach}. The input to this framework is a time-resolved flow-field that is represented on a grid. At each time-step, regions of strain and rotation-dominated flow are identified as grid points which satisfy $Q < -\epsilon_S | Q_{max}|$ and $Q > \epsilon_{\Omega} |Q_{max}|$ thresholds respectively, where $|Q_{max}|$ is representative of the maximum magnitude of $Q$ in the flow. The thresholds $\epsilon_S$ and $\epsilon_{\Omega}$ are chosen to be small numbers, typically of $O(10^{-3})$, so as to provide an accurate identification of strain and rotation dominant regions while being relatively insensitive to numerical errors. In each case, this thresholding yields a subset of the total grid which consists of several disconnected and dense clusters of grid points. These grid points are then segmented into distinct flow structures using the density-based clustering algorithm DBSCAN \citep{Ester1996ANoise}. This is followed by a tracking procedure which allows us to consistently label each segmented flow structure over multiple time-steps. This tracking is based on comparing the location of clusters between time-steps using a simple model of scalar convection. By isolating appropriate spatial volumes corresponding to these structures as they evolve spatio-temporally, we can then rigorously evaluate the forces induced on the airfoil by each of these structures as per equation \ref{eq:VIF}. 

For computational feasibility, we perform this isolation, tracking and force evaluation within a sub-domain of size $2.5C \times 2C$ around the airfoil within which the computed vortex-induced lift is approximately equal to that integrated over the entire flow domain. Furthermore, we partition the vortex-induced lift into components associated with positive $Q$ regions and negative $Q$ regions; these are referred to as ``rotation-induced" and ``strain-induced" lift, respectively, and the corresponding lift coefficients denoted by $C_L^\Omega$ and $C_L^S$, respectively. 

Before the results are presented, it is instructive to place this method within the context of past studies that have quantified the aerodynamic loads induced by individual vortices in a flow. Two studies that are particularly relevant here are those of \cite{PittFord2013LiftVortex} and \cite{Onoue2016VortexPlate} who analyzed the loading induced by the leading and trailing-edge vortices using experimental measurements in unsteady airfoil flows. The common features of the approaches employed by \cite{PittFord2013LiftVortex} and \cite{Onoue2016VortexPlate} are: (a) the assumption of two-dimensionality; (b) use of inviscid models with point vortices in the field combined with conformal mapping and image-vortices to satisfy no-penetration on the airfoil surface; (c) parameterization of vortex circulation from experimental data;  and (d) calculation of induced aerodynamic load via the Blasius formula \citep{BATCHELOR}. The predictions from such models can be quite good and lead to excellent insights into vortex-induced aerodynamic loads.

In comparison, the force partitioning method provides an approach to vortex-induced load quantification that neither requires the assumption of two-dimensionality (see \cite{Zhang2015CentripetalInsects} for application to a 3D flow) nor that of inviscid flow. This eliminates the need for conformal mapping and the difficulties associated flow regularization around sharp edges (by enforcing the Kutta condition). It also removes errors associated with the inviscid flow assumption and the use of point-vortex models. By alleviating these constraints, the force partitioning method demonstrated here is therefore rigorously applicable to general viscous flows. This method allows the evaluation of aerodynamic loading due to individual flow structures in flows that contain several such features interacting and deforming in complicated ways. Furthermore, this method also provides the contribution of acceleration reaction effects as well as viscous diffusion to aerodynamic loads, and can therefore also be applied to flows at low Reynolds numbers. Finally, while the force partitioning is applied here to data from viscous flow simulations, it is, in principle, applicable to experimental data as well.


\section{Results}
\label{sec:results}
\begin{figure}
  \centerline{\includegraphics[scale=1.0]{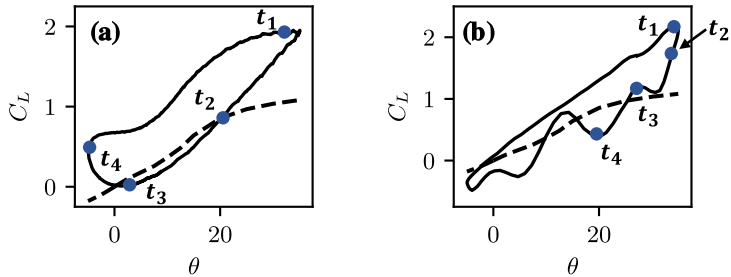}}
  \caption{Coefficient of lift ($C_L$) versus pitch-angle ($\theta$) plots for the two cases: (a) $f = 0.35$, $A_{\theta}=20^{\circ}$; (b) $f = 0.10$, $A_{\theta}=20^{\circ}$. Labels $t_1$--$t_4$ correspond to time-instances examined in subsequent discussion and figures. The dashed line indicates the lift coefficient for corresponding static airfoil \citep{Menon2020AerodynamicNumbers}.}
\label{fig:cl_aoa}
\end{figure}

The analysis is based on two cases of a pitching airfoil with different dimensionless pitching frequencies: $f = 0.35$ and $0.10$. The other parameters which are kept the same for these cases are: $Re=1000$,  $\theta_0 =15^{\circ}$, and $A_{\theta} =20^{\circ}$.
Figure \ref{fig:cl_aoa} shows the lift versus pitch angle for the two cases. While both cases exhibit the classic features of dynamic stall (instantaneous lift that significantly exceeds the corresponding static lift followed by a rapid drop in lift), the differences between the two due to the pitching frequency provide an effective substrate to examine vortex-induced forces in a comprehensive manner. 



\subsection{Case 1: f=0.35}

\begin{figure}
  \centerline{\includegraphics[scale=1.0]{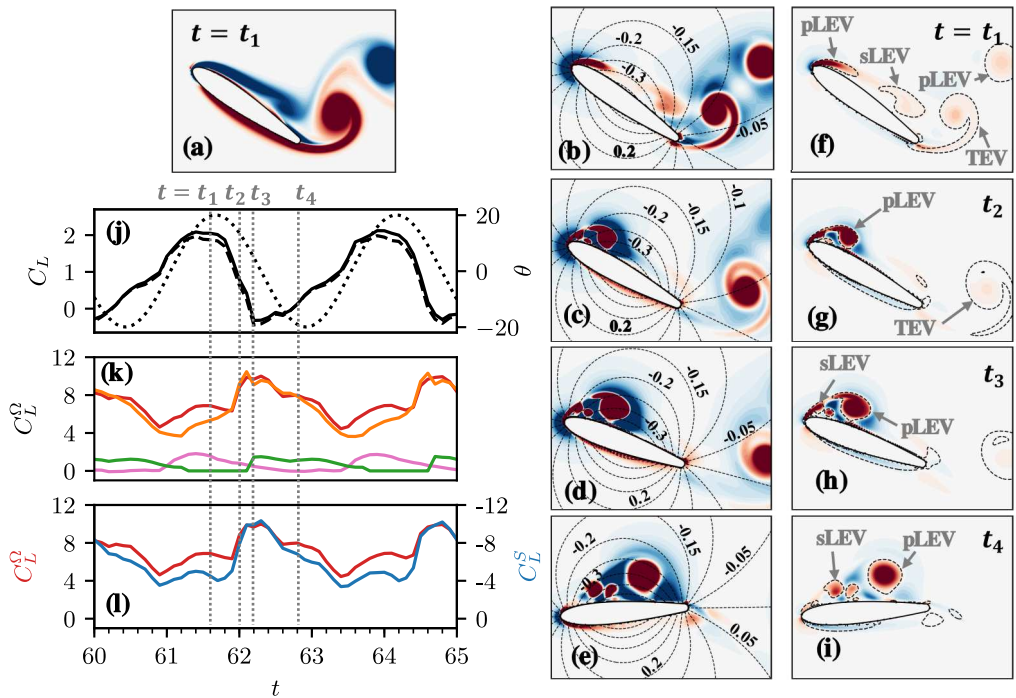}}
  \caption{Flow snapshots and lift contributions for the case with $f=0.35$. (a) Contours of spanwise vorticity at time $t_1$. (b)-(e) Filled contours of $Q$, with levels $[-40,40]$, overlaid with line contours of $\phi_2$. (f)-(i) Distribution of vortex-induced lift ($f_2^\omega = -2 Q \phi_2$) with contour levels $[-65,65]$. Dashed lines and labels show integration volumes corresponding to each vortex which are used to compute the vortex-induced lift associated with these regions as per equation \ref{eq:VIF}. Snapshots (b)-(e) and (f)-(i) are at time-instances $t_1$ -- $t_4$ from top to bottom. (j) Total lift coefficient ($C_L$; \protect\blackline) compared with vortex-induced lift ($F_2^{\omega}$; \protect\blackdashedline). Pitch angle is shown for reference ($\theta$; \protect\blackdottedline). (k) Rotation-induced lift ($C^{\Omega}_L$; \protect\redline) and contributions from the primary LEV (\protect\orangeline), TEV (\protect\pinkline) and secondary LEV (\protect\greenline). (l) Rotation-induced lift due to regions of $Q>0$ ($C^{\Omega}_L$; \protect\redline) and strain-induced lift due to regions of $Q<0$ ($C^{S}_L$; \protect\blueline). Time instances $t_1$ -- $t_4$ are indicated using dashed grey lines.}
\label{fig:cl_snapshots_0.35_20}
\end{figure}
We first discuss the case with $f=0.35$, and specifically examine the time-instances marked $t_1$--$t_4$ in figure \ref{fig:cl_aoa}. Figure \ref{fig:cl_snapshots_0.35_20}(a) shows a snapshot of vorticity at time $t_1$, and we see four key features: a large primary LEV (pLEV) from the previous cycle which is in the near-wake of the foil, a large TEV which is near the trailing edge, a secondary LEV (sLEV) which is located near mid-chord and a shear layer on the suction side which is the initial stage in the formation of the next pLEV.

Figures \ref{fig:cl_snapshots_0.35_20}(b)-(e) show the $Q$ fields at time instances $t_1$ -- $t_4$, overlaid with contours of $\phi_2$. These $Q$ fields clearly identify the key vortices that are generated during the pitching cycle as regions of positive $Q$ and for time $t_1$, figure \ref{fig:cl_snapshots_0.35_20}(b) can be correlated  with the vorticity plot in figure \ref{fig:cl_snapshots_0.35_20}(a). Also noticeable in the plots of $Q$ are the large regions of $Q<0$ that surround the vortices, which corroborates our earlier discussion about the existence of such regions around vortices. These regions of negative $Q$ tend to be more spread out than the vortex cores. There is also a region of negative $Q$ near the leading edge of the airfoil that is sustained throughout the pitching cycle.

The distribution of $\phi_2$ has some distinctive features that have important implications for the vortex-induced lift. First, it has opposite sign over the two surfaces of the airfoil. Thus a vortex adjacent to one surface of the airfoil would generate a force in the opposite direction to that induced by the same vortex adjacent to the other surface. Second, the magnitude of $\phi_2$ is highest near mid-chord and decays towards the leading and trailing edges. Therefore a vortex located near mid-chord would generate a larger lift force than the same vortex located near the leading or trailing-edges. Finally, $\phi_2$ decays with distance from the airfoil and hence, as expected, the influence of a vortex on the lift would diminish with distance from the airfoil.

Figures \ref{fig:cl_snapshots_0.35_20}(f)-(i) show corresponding contour plots of the integrand of the vortex-induced lift, $f_2^\omega = -2 Q \phi_2$. This represents the vortex-induced lift per unit volume of fluid and these plots can be interpreted based on the distribution of $Q$ and $\phi_2$ shown in figures \ref{fig:cl_snapshots_0.35_20}(b)-(e). We see that the most lift-inducing flow structures are those with large $Q$ magnitudes and/or those near mid-chord due to the large magnitude of $\phi_2$. 

The dynamical influence of the various flow structures observed in the flow snapshots discussed above are quantified in figures \ref{fig:cl_snapshots_0.35_20}(j)-(l). Figure \ref{fig:cl_snapshots_0.35_20}(j) shows the time-variation of the total lift coefficient, shown as a solid black line, and the vortex-induced lift coefficient (dashed black line) over two pitching cycles. First, it is noted that the vortex-induced lift is nearly identical to the total lift and this highlights the predominance of this component of the lift force. As mentioned in section \ref{sec:fpm}, the added mass and viscous contributions have orders of magnitude $O(10^{-2})$ and $O(10^{-1})$ respectively. Second, the lift curve shows a rapid drop between $t_1$ and $t_3$, which represents the typical signature of dynamic stall. 
The generally accepted notion is that dynamic stall occurs when an attached leading-edge vortex, which generates high lift, undergoes a ``shedding" event \citep{Carr1988ProgressStall,Dickinson1993UnsteadyNumbers,Chen2010ThePlate}. We can evaluate this notion rigorously by computing the distinct lift contributions of each vortex highlighted above. This is done using an automated framework that segments and tracks the volumes of the $Q>0$ regions occupied by each of the vortices. These spatial volumes are shown as $Q=0$ contours using the dashed lines in figures \ref{fig:cl_snapshots_0.35_20}(f)-(i). The resulting ``rotation-induced lift" for the various vortices is presented in figure \ref{fig:cl_snapshots_0.35_20}(k). We see that the pLEV is indeed the dominant generator of lift on the airfoil, with the sLEV and the TEV generating much smaller contributions (peak lift due to sLEV and TEV is about 15\% of the pLEV). However, very surprisingly, we note that from time $t_1$ to $t_3$ while the total lift on the airfoil falls, the lift induced by the pLEV actually increases and reaches a peak at $t_3$. This increase in pLEV-induced lift is consistent with its growth while staying attached in a region of relatively large $\phi_2$, as seen in figures \ref{fig:cl_snapshots_0.35_20}(c)-(d) as well as \ref{fig:cl_snapshots_0.35_20}(g)-(h). When the pLEV is released and starts to move away from the surface of the foil, as seen in figures \ref{fig:cl_snapshots_0.35_20}(e) and \ref{fig:cl_snapshots_0.35_20}(i) at $t_4$, the lift induced by it reduces, as expected. However, the total lift actually increases during this time, from $t_3$ to $t_4$. Thus, the total lift on the airfoil is nearly perfectly out of phase with the lift induced by the pLEV. This observation runs contrary to our phenomenological understanding of the role of LEVs in lift enhancement and dynamic stall. In fact, this phase-lag between the growth of the pLEV and the lift maximum has been previously reported by \cite{Xia2013LiftPlate} and \cite{Gharali2013DynamicVelocity}.

The cause of this seemingly anomalous behavior becomes clear when we examine the components of the vortex-induced lift associated the negative $Q$ regions (what we call ``strain-induced lift") as shown in figure \ref{fig:cl_snapshots_0.35_20}(l). This plot shows that even though the positive lift induced by the growing LEV does indeed increase between phases $t_2$ and $t_3$, this increase is counteracted by an increase in the negative lift induced by regions of $Q<0$ that surround the LEV. These regions of negative $Q$ occur in areas where the $\phi_2$ has a high magnitude, as seen in figures \ref{fig:cl_snapshots_0.35_20}(c)-(e), and that helps accentuate their effect on the total lift. Thus, the rapid drop in lift occurs not due to the ``shedding" of the LEV but due to the growth of the strain-dominated negative $Q$ regions that develop around the growing LEV. Similarly, the rise in total lift between $t_3$ and $t_4$, which occurs despite the reduction in pLEV-induced lift, is driven by the slightly more rapid reduction in the negative strain-induced lift. Some studies have in fact, reported a phase lag between the onset of dynamic stall and attainment of peak LEV circulation \citep{Xia2013LiftPlate,Gharali2013DynamicVelocity}.
For the current case, we show that such a phase lag, which runs counter to our intuition regarding dynamic stall, can be explained by the significant influence of strain-induced loading. Thus, this case challenges our conventional understanding of phenomenological connection between LEV evolution and dynamic stall, and points to the significant role that the strain dominated regions can play in determining the aerodynamic loads on the surface.

\subsection{Case 2: f=0.10}
\begin{figure}
  \centerline{\includegraphics[scale=1.0]{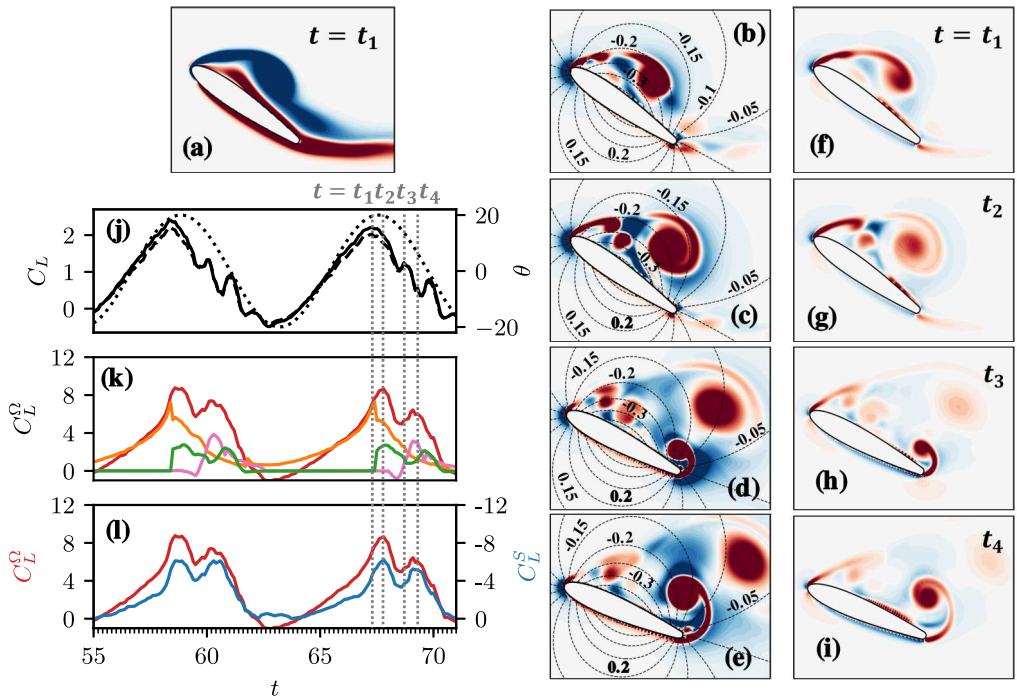}}
  \caption{Flow snapshots and lift contributions for the case with $f=0.10$. (a) Contours of spanwise vorticity at time $t_1$. (b)-(e) Filled contours of $Q$, with levels $[-40,40]$, overlaid with line contours of $\phi_2$. (f)-(i) Distribution of vortex-induced lift ($f_2^\omega = -2 Q \phi_2$) with contour levels $[-65,65]$. Snapshots (b)-(e) and (f)-(i) are at time-instances $t_1$ -- $t_4$ from top to bottom. (j) Total lift coefficient ($C_L$; \protect\blackline) compared with vortex-induced lift ($F_2^{\omega}$; \protect\blackdashedline). Pitch angle is shown for reference ($\theta$; \protect\blackdottedline). (k) Rotation-induced lift ($C^{\Omega}_L$; \protect\redline) and contributions from the primary LEV (\protect\orangeline), TEV (\protect\pinkline) and secondary LEV (\protect\greenline). (l) Rotation-induced lift due to regions of $Q>0$ ($C^{\Omega}_L$; \protect\redline) and strain-induced lift due to regions of $Q<0$ ($C^{S}_L$; \protect\blueline). Time instances $t_1$ -- $t_4$ are indicated using dashed grey lines.}
\label{fig:cl_snapshots_0.10_20}
\end{figure}
We now contrast this behaviour with that at low-frequency oscillations of $f=0.10$. The hysteresis plot in figure \ref{fig:cl_aoa}(b) shows that the pitch-angle reversal in this case corresponds to a sharper initial drop in lift, indicating that the vortices in this low frequency oscillation interacts differently with the timescale of oscillation. The snapshots of $Q$, shown for this case in figures \ref{fig:cl_snapshots_0.10_20}(b)-(e), highlight the main vortex structures in the flow. We see that it is dominated by the same three vortices as in the previous case - the primary and secondary LEVs and the TEV. However, on comparing figures \ref{fig:cl_snapshots_0.10_20}(b) and \ref{fig:cl_snapshots_0.35_20}(d), which correspond to time instances soon after the pLEV-separation in the present and previous case respectively, we see that the pLEV in the current low-frequency case is larger in size and farther away from the surface of the airfoil. This is a result of the slower pitching motion, which allows more time for the growth and downstream motion of the pLEV. 

This increased distance of the pLEV from the surface has important consequences for the lift induced by the pLEV as well as its surrounding region of strain. Figures \ref{fig:cl_snapshots_0.10_20}(f)-(i) show snapshots of $f_2^\omega$ for this case, and we see from figures \ref{fig:cl_snapshots_0.10_20}(g) and \ref{fig:cl_snapshots_0.10_20}(h) that the induced-force density of the pLEV as well as the straining regions around it are significantly smaller than in the $f=0.35$ case discussed earlier (note the difference in $f_2^\omega$ contour levels between the two cases). In particular, a comparison of the lift distribution in figures \ref{fig:cl_snapshots_0.10_20}(g)-(i) with that of figures \ref{fig:cl_snapshots_0.35_20}(g)-(i) shows a stark contrast in the distance of the pLEV from the wall soon after pLEV-separation, as well as the resultant confinement of the strain associated with the pLEV near the wall.

The implications of these observations are evident in figures \ref{fig:cl_snapshots_0.10_20}(j)-(l). Figure \ref{fig:cl_snapshots_0.10_20}(j) again shows that the vortex-induced lift dominates the overall lift production. The contributions of total rotation-induced lift (i.e. $Q>0$ regions) as well as those of the individual vortices are quantified in figure \ref{fig:cl_snapshots_0.10_20}(k). First, we note that the peak in the total rotation-induced lift aligns (at time $t_2$) well with that for the total lift. This is in contrast to the previous case where the the two were quite out-of-phase. Second, the contribution of the pLEV peaks at time $t_1$, which is in-phase with $C_L$ but slightly ahead of the the peak rotation-induced lift. The sLEV appears around this time and augments the rotation-induced lift, resulting in the peak $C_L^{\Omega}$ at $t_2$. There is also a second peak in the total rotation-induced lift at $t_4$ which, as evident from figure \ref{fig:cl_snapshots_0.10_20}(k), is generated by the trailing-edge vortex. 

Finally, figure \ref{fig:cl_snapshots_0.10_20}(l) shows the contribution of the rotation and strain-induced lifts to the total vortex-induced lift. We note that the strain-induced lift has a time-profile that is similar to the rotation-induced lift but remains consistently lower in magnitude for most of the pitching cycle. This is consistent with the observation that regions of large-strain around the vortices are generated relatively further away from the airfoil in this case as compared to the previous case. Furthermore, this also explains the fact that $C_L$ is in-phase with the rotation-induced lift in this case, unlike the behaviour seen for $f=0.35$.

Thus, this lower frequency case presents a more conventional phenomenology of dynamic stall where the total lift correlates, as expected, with the growth and movement of the rotational cores of the vortices. The strain-induced lift force, despite being comparable in magnitude to the rotation-induced lift and of significantly greater magnitude than the total lift, does not play as significant a role for this case, as it does for the first one. 

\section{Conclusions}
\label{sec:conclusions}
We have shown that while studies of vortices and their effects on aerodynamic loads have mostly focused on the rotational cores of vortices, vortex cores are in fact surrounded by dynamically-significant regions where strain dominates over rotation. In particular, these strain-dominated regions exert a considerable, and sometimes even dominant, influence on the induced aerodynamic loads. By using simulations of flow past a pitching airfoil and employing a powerful force partitioning method to separate the force contributions of various regions of the flow, we have shown that the lift induced by the strain dominated regions around vortices is of a magnitude comparable to that induced by the rotational cores of of the vortices. For one of the two cases presented here, we also show that dynamic stall, i.e. the rapid drop in lift for an unsteady airfoil, is less connected with the conventional notion of leading-edge vortex shedding and more with the formation and evolution of the strain dominated ``corona" of the leading edge-vortices.

We close by pointing out that previous investigations \citep{PittFord2013LiftVortex,Xia2013LiftPlate,Wang2013Low-orderFormation,Onoue2016VortexPlate} of vortex-induced aerodynamic loads do implicitly include the effect of strain-dominated regions of a vortex in estimating induced loads. However, the accompanying interpretations of the mechanisms and phenomenology has been based primarily on the rotational vortex cores and has largely ignored the strain-dominated regions associated with vortices. The current study showcases the significant, and sometimes dominant influence of the strain-dominated regions around a vortex, and highlights the utility of explicitly accounting for these effects when examining the mechanisms that govern aerodynamic loading in vortex-dominated flows.

\section*{Acknowledgments}We acknowledge support from AFOSR grant number FA9550-16-1-0404, monitored by Dr. Gregg Abate, and XSEDE computational resources supported by the NSF through allocation TG-CTS100002.

\section*{Declaration of interests}
The authors report no conflict of interest.

\appendix

\bibliographystyle{jfm}
\bibliography{references}

\end{document}